\documentclass[conference]{IEEEtran}

\usepackage[numbers]{natbib}
\usepackage{amssymb}
\usepackage{amsmath}
\usepackage{amsthm}
\usepackage{amsfonts}
\usepackage{tikz}
\usepackage{enumitem}

\usetikzlibrary{arrows}
\usetikzlibrary{decorations.markings}
\usetikzlibrary{shapes}
\tikzset{->-/.style={decoration={
  markings,
  mark=at position .5 with {\arrow{>}}},postaction={decorate}}}
\tikzset{-->-/.style={decoration={
  markings,
  mark=at position .8 with {\arrow{>}}},postaction={decorate}}}

\newcommand{\restr}[1]{\restriction_{#1}}
\newcommand{\dom}{\mathrm{d}}
\newcommand{\s}{\mathcal{S}}

\newcommand{\zer}{\mathbf{0}}
\newcommand{\N}{\mathcal{N}}
\newcommand{\I}{\mathcal{I}}
\newcommand{\R}{\mathcal{R}}

\newcommand{\X}{\mathcal{X}}

\newtheorem{theorem}{Theorem}
\newtheorem{proposition}[theorem]{Proposition}
\newtheorem{lemma}[theorem]{Lemma}
\newtheorem{corollary}[theorem]{Corollary}
\newtheorem{problem}[theorem]{Problem}

\def\set#1{\{#1\}}
\def\c#1{{\mathcal #1}}
\def\M{{\bf M}}
\def\A{{\bf A}}
\def\L{{\sf L}}

\begin{document}
%
\title{Demonic Lattices and Semilattices in Relational Semigroups with Ordinary Composition}

\author{\IEEEauthorblockN{Robin Hirsch}
\IEEEauthorblockA{Department of Computer Science\\University College London\\
Gower St, London WC1E 6EA\\
Email: r.hirsch@ucl.ac.uk}
\and
\IEEEauthorblockN{Ja{\v s} {\v S}emrl}
\IEEEauthorblockA{Department of Computer Science\\University College London\\
Gower St, London WC1E 6EA\\
Email: j.semrl@cs.ucl.ac.uk}
}

\IEEEoverridecommandlockouts
\IEEEpubid{\makebox[\columnwidth]{978-1-6654-4895-6/21/\$31.00~
\copyright2021 IEEE \hfill} \hspace{\columnsep}\makebox[\columnwidth]{ }}
\maketitle
\begin{abstract}
Relation algebra and its reducts provide us with a strong tool for reasoning about nondeterministic programs and their partial correctness. Demonic calculus, introduced to model the behaviour of a machine where the demon is in control of nondeterminism, has also provided us with an extension of that reasoning to total correctness.

We formalise the framework for relational reasoning about total correctness in nondeterministic programs using semigroups with ordinary composition and demonic lattice operations. We show that the class of representable demonic join semigroups is not finitely axiomatisable and that the representation class of demonic meet semigroups does not have the finite representation property for its finite members.  

For lattice semigroups (with composition, demonic join and demonic meet) we show that the representation problem for finite algebras is undecidable, moreover the finite representation problem is also undecidable.  It follows that   the representation class is not finitely axiomatisable, furthermore the finite representation property fails.  
\end{abstract}
\ifCLASSOPTIONpeerreview
\begin{center} \bfseries EDICS Category: 3-BBND \end{center}
\fi
%
\IEEEpeerreviewmaketitle

\section{Introduction}
Binary relations have been used extensively to model nondeterministic programs \cite{mili1987relational, dijkstrapred}. Using the ordinary (or `angelic') composition of binary relations, one can reason about correctness of nondeterministic programs, much  as structures of partial functions can be utilised to reason about correctness of deterministic programs. However, only partial correctness can be deduced when utilising the `angelic' calculus. This is why demonic operations and predicates, analogous to their `angelic' counterparts have been defined to extend this framework to reason about total correctness \cite{dijkstrapred, berghammer1986relational, de2006demonic}.

Relation Algebras are algebras with operators, defined by a finite set of equations, intended to model algebras of binary relations with ordinary identity, converse, composition and Boolean set operations \cite{tarski1941calculus}.  However, not all relation algebras are representable as binary relations, indeed the representation problem for finite relation algebras is undecidable, inferring several further negative computational results.  However, the computational properties of the representation class may improve if we take subsignatures of the relation algebra signature and weaken the definition of representation, accordingly.

We formalise the framework for relational reasoning about total correctness in nondeterministic programs using semigroups with ordinary composition and demonic lattice operations. 
We consider signatures consisting of either or both lattice operators together with composition, where composition  is interpreted angelically but  any semilattice operators are interpreted demonically.
Our three main results:
\begin{description}
    \item[$\set{\sqcup, ;}$] the class of representable demonic join semigroups is not finitely axiomatisable,
    \item[$\set{\sqcap, ;}$] the  class of representable demonic meet semigroups does not have the finite representation property for its finite members,
    \item [$\set{\sqcup,\sqcap, ;}$]  the class of finite representable demonic lattice semigroups is not recursive, from which non-finite axiomatisability and failure of the finite representation property follow.  
\end{description}

\subsection{Preliminaries}

When representing a first-order structure as a concrete structure of binary relations, we can interpret the lattice operations/predicate  $+,\cdot, \leq$  as $\cup, \cap, \subseteq$ respectively. However, one can also define demonic join ($\sqcup$) for $R,S \subseteq X \times X$ as
\begin{equation}\label{eq:sqcup}R \sqcup S = (R \cup S)\restr{\dom(R) \cap \dom(S)}\end{equation}
where $R\restr{\dom(S)}$ denotes the restriction of relation $R$ to $\dom(S)$, i.e.
\begin{gather*}
    R\restr{\dom(S)} = \{(x,y) \in R \mid x \in \dom(S)\}\\
    \dom(S) = \{x \in X \mid \exists y \in X: (x,y) \in S\}.
\end{gather*}

Demonic join defines demonic refinement analogous to their `angelic' counterparts. Specifically, $R \subseteq S \Longleftrightarrow R \cup S = S$ and $R \sqsubseteq S \Longleftrightarrow R \sqcup S = S$. Concretely,
\begin{equation}\label{eq:sqsub}R \sqsubseteq S \Longleftrightarrow \bigg(\dom(S) \subseteq \dom(R) \wedge R\restr{\dom(S)} \subseteq S\bigg).\end{equation}

For demonic meet we appear to hit a problem, since two binary relations are not sure to have any common demonic refinement.   A necessary and sufficient condition for two relations $R, S$ to have a common demonic refinement is 
\begin{equation}\label{eq:27}
\dom(R)\cap \dom(S)=\dom(R\cap S).
\end{equation}
For the necessity of this condition, suppose $T\sqsubseteq R, S$ and  $x\in \dom(R)\cap \dom(S)\subseteq \dom(R)\subseteq \dom(T)$.  Then there is $y$ such that $(x, y)\in T$.  Since $x\in \dom(R), \dom(S)$ and $T\sqsubseteq R, S$, we get $(x, y)\in R$ and $(x, y)\in S$, hence $(x, y)\in (R\cap S)$,\/ $x\in \dom(R\cap S)$.  Thus $T\sqsubseteq R, S$ implies $\dom(R)\cap \dom(S)\subseteq \dom(R\cap S)$, the reverse inclusion is trivial.  Conversely, given \eqref{eq:27}, a common demonic refinement of $R, S$ is given by

\begin{equation}\label{eq:sqmeet}
    R\sqcap S =  (R\cap S)\cup R\restr{\setminus \dom(S)}\cup S\restr{\setminus \dom(R)}.
\end{equation}
where $\setminus \dom(s)$ denotes the complement of the domain of $s$.

Given an arbitrary set $\c S$ of binary relations over the base $X$,  we may extend the base to $X'=X\cup\set\times$ and replace each relation $R\in\c S$ by the relation $R'=R\cup\set{(x,\times):x\in d(R)}$ and let $\c S'=\set{R':R\in\c S}$.   Observe that the map $R\mapsto R'$ is a bijection from $\s$ to $\s'$, preserving unions, intersections, compositions and domains.  Moreover, for all $R', S'\in\c S'$ we have $\dom(R')\cap \dom(S')=\dom(R'\cap S')$, so their demonic meet    may be defined by \eqref{eq:sqmeet}.

The abstract composition operation ($\circ$) can also be interpreted as ordinary composition ($;$) and as demonic composition ($*$) defined for some relations $R,S$ over $X$ as
\begin{gather}
\label{eq:comp}    R;S = \{(x,z) \in X^2 \mid \exists y ( (x,y) \in R \wedge (y,z) \in S)\}\\
  \nonumber  R*S = \{(x,y) \in R;S \mid \forall z  (x,z) \in R \Rightarrow z \in \dom(S)\}
\end{gather} 

We focus on  the three signatures $\tau$  where $\set{\sqsubseteq,;}\subseteq \tau\subseteq\set{\sqsubseteq,\sqcup, \sqcap,;}$.
A $\tau$-structure $\c S$  is just a set with a binary relation for $\sqsubseteq$ and binary (partial) functions for $;, \sqcup, \sqcap$ if in $\tau$.  If there is a set $X$ (the base),   each $s\in\c S$ is a binary relation over $X$, and operations are defined set-theoretically, by  \eqref{eq:sqcup}, \eqref{eq:sqsub}, \eqref{eq:sqmeet}, \eqref{eq:comp},  for $\sqcup,\sqsubseteq,  \sqcap,;$ respectively, then $\c S$ is \emph{proper}.    An isomorphism from a $\tau$-structure to a proper $\tau$-structure is called a \emph{$\tau$-representation}.

The class of all $\tau$-representable structures is called a representation class, denoted $R(\tau)$. $R(\tau)$ is finitely axiomatisable if and only if a finite $\tau$ theory $\Psi$ exists such that $\s \models \Psi \Longleftrightarrow \s \in R(\tau)$.

For a signature $\tau$ we say that $R(\tau)$ has the \emph{finite representation property} iff every finite $\tau$-structure in $R(\tau)$ has a $\tau$-representation over a finite base.

These properties provide computing guarantees for reasoning about these structures. Finite axiomatisability implies that the representability decision problem is in $LOGSPACE$ and the finite representation property implies that it is decidable. Furthermore, an explicit definition of a representation over a finite base can improve the upper bound on the complexity of that problem as well as provide a framework of switching between reasoning about the concrete and abstract structures.

\subsection{Related Work}

The representation class $R(\subseteq, ;)$ is a great example of good behaviour in reducts of relation algebra. Zarecki{\u \i}'s axioms for ordered semigroups \cite{zareckii1959representation} axiomatises the class by associativity, partial order and monotonicity. A simple finite representation for finite structures is also provided. A $(\leq, \circ)$-structure $\s$ is amended with an element $e$ such that $\forall s: s \circ e = e \circ s = s$ and $\forall s: (s \leq e \vee e \leq s) \Rightarrow s = e$ and a mapping $\theta: \s \rightarrow \wp((\s \cup \{e\})^2)$  where
$$(s,t) \in a^{\theta} \Longleftrightarrow s \leq a \circ t$$

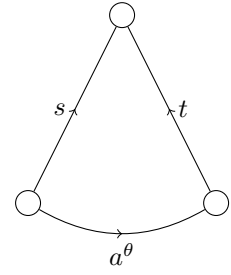
\begin{figure}
    \centering
    \begin{tikzpicture}
        \coordinate[draw,circle] (s1) at (0,0);
        \coordinate[draw,circle] (t1) at (2.5,0);
        \coordinate[draw,circle] (o1) at (1.25,2.5);

        \path(s1) edge[->-, bend right] node[below]{$a^{\theta}$} (t1);
        
        \path(s1)edge[->-] node[left]{$s$} (o1);
        
        \path(t1)edge[->-] node[right]{$t$} (o1);
    \end{tikzpicture}
    \caption{Zarecki{\u \i}'s Representation $\theta$ for ordered semigroups}
    \label{fig:dualZareckii}
\end{figure}

The mapping can be visualised as in Figure~\ref{fig:dualZareckii}.

It turns out that the axioms of ordered semigrouos also define $R(\sqsubseteq, *)$. However, recently, it has been shown that there is no finite axiomatisation for $R(\subseteq, *)$ \cite{hirsch2020algebra} or $R(\sqsubseteq, ;)$ \cite{hirsch2020finite}. The latter citation shows that  $R(\sqsubseteq, ;)$ has the finite representation property,  partly based on Zarecki{\u \i}'s $\theta$, above. For $R(\cup, ;), R(\sqcup, *)$, it remains unknown whether the finite representation property holds or not.

A number of results in the field suggests that the good behaviour of the ordered semigroups does not extend to the rest of the angelic signatures. $R(\cap, ;)$ is finitely axiomatisable, but not finitely representable \cite{bredihin1978representations, neuzerling2016undecidability}, $R(\cup, ;)$ is not finitely axiomatisable \cite{andreka1988representation} and the representation problem for $R(\cup,\cap, ;)$ is undecidable \cite{neuzerling2016undecidability}, implying both non-finite-axiomatisability and failure of finite representation property. These results are summarised in Table~\ref{tab:results}, along with the contributions of this paper.

\begin{table}[b]
    \centering
    \begin{tabular}{|c|c|c|}
        \hline
        $\tau$ & $R(\tau)$ FA & FRP \\ \hline \hline
        $\{\subseteq, ;\}$ & \checkmark \cite{zareckii1959representation} & \checkmark \cite{zareckii1959representation} \\
        $\{\sqsubseteq, ;\}$ & $\times$ \cite{hirsch2020finite} & \checkmark \cite{hirsch2020algebra} \\ \hline
        $\set{\cup, ;}$ & $\times$ \cite{andreka1988representation} & \\
        $\set{\sqcup, ;}$ & $\times$ &  \\ \hline
        $\set{\cap, ;}$ & \checkmark \cite{bredihin1978representations} & $\times$ \cite{neuzerling2016undecidability}\\
        $\set{\sqcap, ;}$ & & $\times$ \\ \hline
          $\set{\cup,\cap, ;}$ & $\times$ \cite{neuzerling2016undecidability} & $\times$ \cite{neuzerling2016undecidability}\\
        $\set{\sqcup,\sqcap, ;}$ & $\times$ & $\times$ \\ \hline
    \end{tabular}
    \vspace{0.2cm}
    \caption{Summary of results for `angelic' and demonic lattice semigroups with `angelic' composition}
    \label{tab:results}
\end{table}

\section{Relational Modelling of Termination and Correctness}
\label{sec:mod}

In this section we formalise the framework for relational reasoning about termination and total correctness of nondeterministic programs using demonic lattice and ordinary composition. This is based on the Refinement Algebra \cite{von2004towards} (an extension of Kleene Algebra), but adapted for relational reasoning about correctness as described in \cite{mili1987relational}. 

The terminology and notation we use  is common but not universal. In Kleene Algebraic literature, $\sqsupseteq, \sqcap, \sqcup$ is sometimes used in place of $\sqsubseteq, \sqcup, \sqcap$ respectively, to more closely reflect the behaviour of tests.

A deterministic program can be modelled as a (partial) function from the space of configurations $C$ of the machine to itself where the image of $c_1 \in C$ is the $c_2 \in C$ where the program terminates after its execution starts from $c_1$ if such $c_2$ exists and undefined otherwise. For nondeterministic programs this $c_2$ may not be unique, thus we model a nondeterministic program as a relation.

To define it formally, given the configuration space $C$, we say that a nondeterministic program $A$ is modelled as a relation over the base $C$ where $(c_1,c_2) \in A$ if and only if there exists a possible run of $A$ from $c_1$ that terminates in $c_2$.

Using this type of modelling, we can use relational calculus to model behaviour of programs. Some examples include

\begin{itemize}
    \item Relational composition ($;$) to model sequential runs of two programs
    \item Join ($\cup$) as nondeterministic choice between two programs
    \item Empty relation ($0$) as  either an aborting or as an infinitely looping program.
\end{itemize}

Furthermore,  sub-identity relations over the same configuration space may be used to model conditions, for tests. We say that a condition $P$ is modelled by the set of all pairs $(c,c) \in C \times C$ such that $c \models P$.

Knowing how to relationally model programs and conditions, we are now equipped to talk about correctness. A \emph{Hoare triple} $(P,A,Q)$ consists of a \emph{precondition} $P$, a program $A$ and a \emph{postcondition} $Q$. If $P$ holding prior to the execution is sufficient for any terminating run of $A$ to establish $Q$, we say that $(P,A,Q)$ is \emph{partially correct}. A triple is said to be \emph{totally correct} if it is partially correct and it has a terminating run, assuming the precondition holds prior.

We model partial correctness of a Hoare Triple $(P,A,Q)$ in two ways:
\begin{gather*}
    P;A;Q = P;A \\
    P;A;(\neg Q) = P;0
\end{gather*}

One can easily check that since $0$ is the empty relation and the bottom element of the ordinary lattice, the two equations are equivalent. This will not be the case, as we will see, when we define the two statements in terms of the demonic lattice.

\begin{figure}
    \centering
    \begin{tikzpicture}
        \node[circle, draw] (c1a) at (0,0) {$c_1$};
        \node[circle, draw] (c2a) at (2.5,1) {$c_2$};
        \node[circle, draw] (c3a) at (4,0) {$c_3$};
        \node[circle, draw] (c4a) at (1.5,2) {$c_4$};
        \coordinate (c5a) at (3,2.5);
        
        \path (c1a) edge[->-] node[above]{$A$} (c2a);
        \path (c2a) edge[->-] node[above]{$B$} (c3a);
        \path (c1a) edge[->-] node[above left]{$A$} (c4a);
        \path (c4a) edge[dashed, ->-] node[above]{$B$} (c5a);
        \path (c1a) edge[dashed, bend left=90, ->-] node[above left]{$A*B$} (c5a);
    \end{tikzpicture}
    \begin{tikzpicture}
        \node[circle, draw] (c1a) at (0,0) {$c_1$};
        \node[circle, draw] (c2a) at (2.5,1) {$c_2$};
        \node[circle, draw] (c3a) at (4,0) {$c_3$};
        \node[circle, draw] (c4a) at (1.5,2) {$c_4$};
        \node[circle, draw] (c5a) at (3,2.5) {$c_5$};
        
        \path (c1a) edge[->-] node[above]{$A$} (c2a);
        \path (c1a) edge[->-] node[above]{$A*B$} (c3a);
        \path (c2a) edge[->-] node[above]{$B$} (c3a);
        \path (c1a) edge[->-] node[above left]{$A$} (c4a);
        \path (c4a) edge[->-] node[above]{$B$} (c5a);
    \end{tikzpicture}
    \caption{Motivation for Demonic Composition}
    \label{fig:demComMot}
\end{figure}
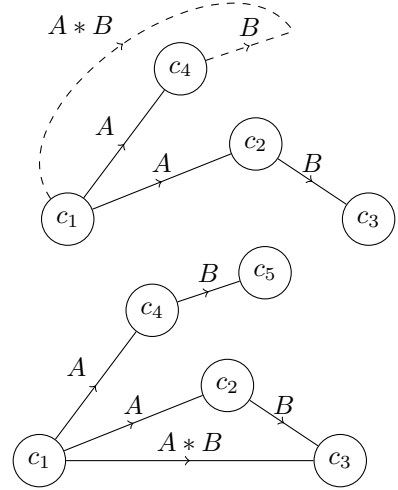

Let us first motivate some demonic calculus operations and constants. We begin with the demonic composition. Imagine the demon was in control of the nondeterminism in the machine. His motivation is to abort or loop indefinitely if possible, otherwise maximise the opportunities to establish the wrong postcondition. So given two programs $A,B$ modelled as binary relations and some $(c_1, c_2) \in A, (c_2, c_3) \in B$, the demon would not include $(c_1, c_3) \in A*B$ if there existed some $c_4, (c_1,c_4) \in A$ with no $c_5$ such that $(c_4, c_5) \in B$ (see top of Figure~\ref{fig:demComMot}). This is because running $B$ from $c_4$ results in abort/infinite loop and so he picks the run of $A$ that takes it to $c_4$ when it is succeeded by $B$, resulting in abort/infinite loop. Alternatively, if for all $c_4, (c_1,c_4) \in A$ there exists a $c_5$ such that $(c_4, c_5) \in B$ (see bottom of Figure~\ref{fig:demComMot}), then $(c_1, c_3)$ should be included in $A*B$ as a run to $c_3$ may establish an undesirable postcondition other runs do not.

Similarly we motivate demonic join with choice in the demonic nondeterminism. Consider again two programs $A,B$ and, without loss, for a pair of  configurations $(c_1,c_2) \in A$ we have $(c_1,c_2) \in A \cup B$. However, if the demon is in control of the machine and there is no $c_3$ such that $(c_1,c_3) \in B$, this means that $B$ aborts or loops infinitely from $c_1$. Thus the demon will choose to run $B$ from $c_1$ and hence $(c_1,c_3) \not \in A \sqcup B$. Otherwise, if $c_1$ is in the domain of $B$, the demon will try to maximise the possibility of the program potentially reaching an undesirable postcondition, so  $(c_1,c_2) \in A \sqcup B$.

The definition of the demonic join enables us to talk about the demonic semilattice. It also enables us to talk about the demonic refinement $\sqsubseteq$. It becomes apparent that the empty relation $0$ is the top element of this lattice. Intuitively speaking, it is easy to see why. As discussed $0$ models the aborting/non-terminating programs. Thus for any program $A$, when the demon is given a choice between $0$ and $A$, he will choose $0$, i.e. $A \sqcup 0 = 0, \;A \sqsubseteq 0$.

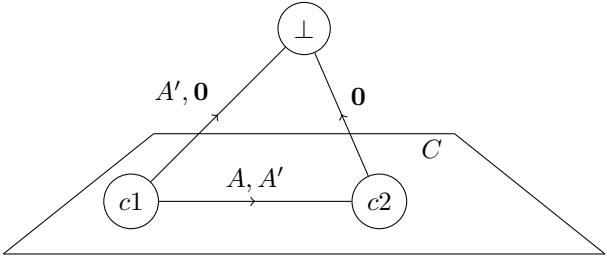
\begin{figure}
    \centering
    \begin{tikzpicture}
        \node[circle, draw] (c1) at (-0.3, 0.2) {$c1$};
        \node[circle, draw] (c2) at (3, 0.2) {$c2$};
        \node[circle, draw] (f) at (2,2.5) {$\bot$};
        
        \coordinate (a) at (-2,-0.5);
        \coordinate (b) at (6,-0.5);
        \coordinate (d) at (0,1.1);
        \coordinate (c) at (4,1.1);
        \draw (a) -- (b);
        \draw (c) -- (b);
        \draw (a) -- (d);
        \draw (c) -- (d);
        \node at (3.7,0.9) {$C$};
        
        \path (c1) edge[->-] node[above]{$A, A'$} (c2);
        \path (c1) edge[->-] node[above left]{$A', \zer$} (f);
        \path (c2) edge[->-] node[above right]{$\zer$} (f);
    \end{tikzpicture}
    \caption{Adding the `magic' configuration to our model of programs}
    \label{fig:addMagic}
\end{figure}

For demonic meet,  we introduce a program $\zer$ (note the bold font to distinguish from $0$ above), that the demon would avoid at all costs. In fact, this program is akin to magic, it should both terminate and establish any postcondition, even $\bot$. To model this, we extend the space $C$ to $C'=C\cup\set{\bot}$ and let $$\zer = \{(c, \bot) \mid c \in C\}.$$
For all conditions $P$, including contradictions, we let $P' = P \cup \{(\bot, \bot)\}$ and for all programs $A' = A \sqcup \zer = A \cup \{(c,\bot) \mid c \in \dom(A)\}$, see Figure~\ref{fig:addMagic}.

Observe how $(P', A', Q')$ is still correct if and only if $(P,A,Q)$ is correct - both partially and totally. Intuitively, this is because if the program $A'$ is defined as $A \sqcup \zer$, i.e. the demon's choice between $A$ and `magic', the behaviour of the program will not change. However, observe that the following equations for characterising correctness of $(P',A',Q')$ are no longer equivalent.
\begin{gather*}
    P';A';Q' = P';A'\\
    P';A';(\neg Q') = P';\zer
\end{gather*}

In fact, the former still expresses partial correctness, i.e. all terminating runs of $A'$ from $P'$ establish the postcondition $Q'$. However, the latter now characterises total correctness, i.e. all terminating runs of $A'$ from $P'$ establish $Q'$ and there exists a terminating run of $A'$ from $P'$.

\section{Semigroups with Demonic Join}
\label{sec:join}
In this section we look at the signature $(\sqcup, ;)$, analogous to the abstract $(+, \circ)$.     

The representation class $R(\cup, ;)$ of the corresponding angelic signature was proven non-finitely axiomatisable in \cite{andreka1988representation, andreka2011axiomatizability}, however it remains unknown whether finite algebras in $R(\cup, ;)$ possess finite representations. \cite{andreka2011axiomatizability} shows non-finite axiomatisability by explicitly defining a collection of non-$\set{\cup, ;}$-representable structures with a representable ultraproduct. However, these turn out to be $\set{\sqcup, ;}$-representable, due to domain restriction in the definition of $\sqcup$.

We use representation games based on \cite{hirsch2002relation} to show that although the representation class $R(\sqcup, ;)$ has no finite axiomatisation. The structures used in the argument show affinity for those used in \cite{andreka2011axiomatizability}, however, significant modifications were needed to ensure appropriate domain restrictions and inclusions.

We begin the argument by defining the game, played over networks. A network $\N = (N,\top, \bot)$ is defined for a structure $\s$ as a set of nodes $N$, an edge labelling function $\top: (N \times N) \rightarrow \wp(\s)$ and  a node labelling function $\bot:N\rightarrow\wp(\s)$ (identifying elements forbidden on outgoing edges).
If $\N'=(N', \top', \bot')$ is another network,  we write $\N\subseteq\N'$ and say $\N'$ \emph{extends} $\N$ if $N\subseteq N'$ and for all $x, y\in N$ we have $\top(x, y)\subseteq\top'(x, y),\;\bot(x)\subseteq\bot'(x)$.  

We say $\N$ is \emph{consistent} if and only if, for all $x, y\in N$
\begin{gather*}
\top(x,y) \cap \bot(x) = \emptyset,
\end{gather*}
\emph{closed} if for all $x, y, z\in\c N$,
\begin{gather*}
(a\in\top(x, y)\wedge b\in\c S)\rightarrow ((a+b)\in\top(x, y)\vee b\in\bot(x))\\
 (a\in\top(x, y)\wedge b\in\top(y, z))\rightarrow (a\circ b)\in \top(x, z),
\end{gather*}
and  \emph{saturated} if it is consistent, closed, and for all $x, y\in N$
\begin{align*}
    (a+b)\in\top(x, y)\rightarrow&(a\in\top(x, y)\vee b\in\top(x, y))\\ 
    & \wedge \exists z (a \in \top(x,z) \wedge  \exists z(b \in \top(x,z))\\
    (a\circ b)\in \top(x, y)\rightarrow&\exists z (a\in\top(x, z)\wedge b\in\top(z, y)). 
\end{align*}

\begin{proposition}\label{prop:sat}
If $\N$ is saturated then  the map $a\mapsto\set{(x, y):x, y\in\N,\; a\in \top(x, y)}$ is a homomorphism from $\s$ into $(\wp(N\times N), \sqcup, ;)$.  
\end{proposition}


The game  $\Gamma_n(\s)$ has $n\leq\omega$ moves excluding initialisation (the zeroth move),  played by two players,  Abelard ($\forall$) and Eloise ($\exists$).
A play of the game consists of a sequence of networks $\N_0\subseteq\N_1\subseteq\ldots \N_n$ (for $n<\omega$) or an infinite sequence when $n=\omega$, together with the initial nodes $x_0, y_0\in\N_0$ and  forbidden label $s_\bot\in\s$.

 At the $i$th move, $\exists$ returns a  network $\N_i$ based on what $\forall$ demands.
  We sketch the moves in outline first.  In the \emph{initial} move $\forall$ picks $a\neq b\in\c S$ and demands a  \emph{discriminating}  network $\c N_0$ and $x_0, y_0\in\N_0$, where $\top(x_0, y_0)$ includes one of $\set{a, b}$ and where the forbidden label $s_\bot$ is the other, at $\exists$'s choice.   For a \emph{choice} move $\forall$ picks $(a+b)\in \c N(x, y)$ and $\exists$ must add either $a$ or $b$ to $\top(x, y)$, without loss say $a$,  but also she must ensure that $x$ is in the domain of the other  element $b$ by including a node $z$ (possibly new) where $b\in \top(x, z)$.  For a \emph{witness} move $\forall$ picks $a\circ b\in\top(x, y)$ and $\exists$ must include a node $z$ (possibly new) such that $a\in\top(x, z), \;b\in\top(z, y)$.    \emph{Join} and \emph{composition} moves are designed to ensure the two closure properties.  $\forall$ wins if an inconsistent network is played in any round, or if $s_\bot\in \top(x_0, y_0)$ occurs in any round.  

For the formal definition, we introduce some notation to allow us to add edges and labels to a  network.   For some $\N = (N,\top,\bot)$,\/ $s \in \s$, \/$x \in N$,\/ $y \in N$ we let
\begin{gather*}
    \N^{+_\top}[s,x,y] = (N, \top', \bot)\\
    \N^{+_\bot}[s,x] = (N, \top, \bot')
\end{gather*}
where $\top'$ is identical to $\top$ except $\top'(x, y)=\top(x, y)\cup\set s$ and $\bot'$ is identical to $\bot$ except $\bot'(x)=\bot(x)\cup\set s$.  If $y^+\not\in N$ let $\N^{+_\top}[s, x, y^+]=(N\cup\set {y^+}, \top', \bot, x_0, y_0, \s_\bot)$ where $\top'(x, y)=\top(x, y)$ for $x, y\in N$ and  $\top'(x, y^+)=\set s$,\/ $\top'(x', y^+)=\emptyset$ for $x'\in (N\cup\set{y^+})\setminus\set{x}$.

For initialisation, let  $\N[a, b]$ be the  network with  two distinct nodes $\set{x_0, y_0}$, 
where $\top(x_0, y_0)=\set a$, but all other edges and nodes have empty labels, and forbidden label $s_\bot=b$.  

Now, we have all the tools to define the rules of the game. 

\paragraph{Initialisation Move} $\forall$ picks $a \neq b \in \s$ and $\exists$ plays $x_0, y_0$  in a  network $\N_0$ extending $\N[a,b]$ or $\N[b,a]$.

\paragraph{Choice Move} $\forall$ picks a pair of nodes $x,y$ in the current   network $\N_i$ and an $a + b \in \top(x,y)$.  $\exists$ must pick a node $z$ (existing or new) and return $\N_{i+1}$  extending $\left(\N_i^{+_\top}[x,y,a]\right)^{+_\top}[x,z,b]$ or $\left(\N_i^{+_\top}[x,y,b]\right)^{+_\top}[x,z,a]$.

\paragraph{Join Move} $\forall$ picks a pair of nodes $x,y$ in the current  network $\N_i$,  and some $a \in \top(x,y)$ as well as some $b \in \s$. $\exists$ has a choice of returning an extension of  $\N_i^{+_\top}[x,y,a + b]$ or $\N_i^{+_\bot}[x,b]$.

\paragraph{Witness Move} $\forall$ picks a pair of nodes $x,y\in\c N_i$ and some $a\circ b \in \top(x,y)$. $\exists$ picks a (potentially new) node $z$ and returns an extension of  $\N_{i+1}=\left(\N_i^{+_\top} [x, z, a]\right)^{+_\top} [z, y, b]$.

\paragraph{Composition Move} $\forall$ picks some nodes $x,y,z\in\N_i$ with some $a \in \top(x,y), \; b \in \top(y,z)$ and $\exists$ must respond with an extension of $\N_i^{+_\top}[ x,z,a \circ b]$.

Let $n<\omega$ be finite.  
If after $n$ moves, the  network $\N_n$ is still consistent and $s_\bot\not\in\top_n(x_0, y_0)$, we say $\exists$ has won $\Gamma_n(\s)$. Otherwise, $\forall$ has won.   For the infinite game $\Gamma_\omega(\s)$, if an inconsistent network is played or $s_\bot\in\top(x_0, y_0)$ occurs in any round then $\forall$ has won, if this never happens then $\exists$ is the winner.  For a  network $\N$,\/ $x_0, y_0\in\N,\;s_\bot\in\s$ and $n\leq\omega$ we define a variant $\Gamma_n(\s, \N, x_0, y_0, s_\bot)$ of this game where $\N$ is played in the initial round, and subsequent moves are as above.   

For each $\forall$-move, $\exists$ may play any extension of a finite number of possible networks.  However, if she has a winning strategy in this  game, she will also have a winning strategy where she plays  \emph{conservatively},  by playing one of the alternative networks but not their proper extensions.

\begin{lemma}
\label{lem:repifstrat}
For countable $(+,\circ)$-structures $\s$,  the following are equivalent: (i) $\exists$ has a winning strategy for $\Gamma_n(\s)$ (all $n<\omega$),  (ii) $\exists$ has a winning strategy for $\Gamma_\omega(\s)$, (iii)  $\s$ is $(\sqcup, ;)$-representable.
\end{lemma}

\begin{proof}
(ii) $\Rightarrow$ (i) is trivial.    The implication (i) $\Rightarrow$ (ii) holds, by K\"onig's tree lemma, because if  $\exists$ plays conservatively, she has essentially only finitely many choices for each move.  
To prove that (iii) implies (ii), in a play of $\Gamma_\omega(\s)$,\/  $\exists$ maintains a label preserving map from the nodes of the current  network to the base of the representation, clearly a winning strategy.  

For (ii) $\Rightarrow$ (iii), observe that $\forall$ can schedule his  $\omega$ moves in such a way that every legal move is eventually played (since $\s$ is countable). 
Thus, for $a\neq b\in\c S$ the  limit  network $\c N_\infty[a,b]$ of such a play of the game, where $\exists$ uses her winning strategy and  where the initial move is $(a, b)$,  has a pair of nodes $(x_0, y_0)$ that have either $a$, but no $b$ or $b$ but no $a$ in  
$\top(x_0, y_0)$ (due to initialisation) and is saturated. 
Take  the disjoint union $\c N_\infty= \bigcup_{a \neq b \in S} \N_\infty[a,b]$  by renaming the nodes of each $\c N[a, b]$.  Now the map $\theta:\c S\rightarrow \wp(N_\infty\times N_\infty)$ defined  by  $a^\phi=\set{(x, y): x, y\in N_\infty,\; a\in\top(x,y)}$ is a homomorphism, by Proposition~\ref{prop:sat} discriminating every pair $a\neq b\in \s$, hence a $(\sqcup, ;)$-representation of $\s$.  
\end{proof}

\begin{lemma}
\label{lem:sig}  Let $n<\omega$.  
There exists a formula $\sigma_n$ such that $\s \models \sigma_n$ if and only if $\exists$ has a winning strategy for $\Gamma_n(\s)$. Thus $\set{ \sigma_n: n<\omega}$ axiomatises $R(\sqcup, ;)$.
\end{lemma}

\begin{proof}
Here we consider \emph{term networks}, similar to  networks except the labels are sets of terms built with $\circ, +$ from variables,  rather than elements of $\s$.   A term network $\N$ together with a variable assignment $v:vars\rightarrow\s$ determines a  network $v(\N)$.  For each term network $\N$ with nodes $x_0, y_0$, the set of terms $\bigcup_{x, y\in\N}(\top(x, y)\cup \bot(x))$, and a variable $s_\bot$ we can define $\phi_0(\N)$ by

\begingroup
\allowdisplaybreaks
\[
 \bigwedge_{s \in \top(x_0,y_0)} s \neq s_\bot 
    \wedge \bigwedge_{x,y \in N,\;s \in \top(x,y),\; t \in \bot(x) }  s \neq t 
\]
asserting consistency. Recursively,  $ \phi_{n+1}[\N] $ is the conjunction over all $x, y, z\in N$  of the conjunction of the following four formulas, corresponding to the four types of move.
\begin{description}
\item[Choice]
    \begin{gather*}
\!\!\!\!\!\!\! \bigwedge_{s\in\top(x, y)}\forall u\forall t (s=t+u\rightarrow  \bigvee_{z \in N\cup\{z^+\}}\hspace{1.3in}\;\\
 \!\!\!\!\!\!\!\!\!\!  \phi_n(\left(\N_n^{+_\top}[x,y,t]\right)^{+_\top}[x,z,u])\vee \phi_n(\left(\N_n^{+_\top}[x,y,u]\right)^{+_\top}[x,z,t]))
    \end{gather*}
    
    \item[Join]
 \[ \bigwedge_{s\in\top(x, y)}\forall t(\phi_n(\N^{+_\top}[x,y,s+t]) \vee \phi_n(\N^{+_\bot}[x, t]))\]
    
\item [Witness]
\begin{gather*}   \bigwedge_{s\in\top(x, y)} \forall t,u\;(s = t \circ u \rightarrow \hspace{1in} \; \\ \bigvee_{z \in N \cup \{z^+\}}
      \phi_n((\N^{+_\top}[ x, z, t])^{+_\top}[z, y, u])) \end{gather*}
      
      \item[Composition]
\[    \bigwedge_{s\in\top(x, y),\; t \in \top(y,z)}  \phi_n(\N^{+_\top}[x,z,s\circ t])
\]
\end{description}
\endgroup
It follows directly from the definition of the game and the formula that $\exists$ has a winning strategy for $\Gamma_n(\s, v(\N))$ (where $v(\N), x_0, y_0, v(s_\bot)$ is played in the initial round)  if and only if $\s, v \models \phi_n(\N)$. 
Now we define the sentence
\begin{gather*}
    \sigma_n = \forall a, b\;\Bigg( a \neq b \rightarrow (\phi_n(\N[a,b]) \vee \phi_n(\N[b,a])\Bigg)
\end{gather*}
Then $\exists$ has a winning strategy in $\Gamma_n(\s)$ if and only if $\s\models\sigma_n$. By Lemma~\ref{lem:repifstrat}, a countable structure belongs to $R(\sqcup, ;)$ if and only if it satisfies $\set{\sigma_n:n<\omega}$.  Since the representation class is pseudo-elementary, it is closed under elementary equivalence.  Hence, for an arbitrary structure $\c S$ we can take a countable elementary substructure $\c S_0$ and we have $\c S\in R(\sqcup, ;)$ iff $\c S_0\in R(\sqcup, ;)$ iff $\c S_0\models\set{\sigma_n:
n<\omega}$ iff $\c S\models\set{\sigma_n:n<\omega}$.

\end{proof}

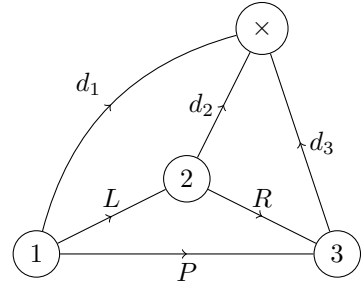
\begin{figure}
    \centering
    \begin{tikzpicture}
        \node[draw, circle] (x) at (0,0) {$1$} ;
        \node[draw, circle] (y) at (2,1) {$2$};
        \node[draw, circle] (z) at (4,0){$3$};
        \node[draw, circle] (f) at (3,3){$\times$};
        
        \path (x) edge[->-] node[below]{$P$} (z);
        \path (x) edge[->-] node[above]{$L$} (y);
        \path (y) edge[->-] node[above]{$R$} (z);
        \path (y) edge[->-] node[left]{$d_2$} (f);
        \path (z) edge[->-] node[right]{$d_3$} (f);
        \path (x) edge[->-, bend left] node[above left]{$d_1$} (f);
    \end{tikzpicture}
    \caption{Domains of $L,R,P,D$}
    \label{fig:domainsSn}
\end{figure}

Now, we define for each $1 \leq n < \omega$ a structure $S_n$ that is not representable, but $\exists$ has a winning strategy for at least $\Gamma_n(\s_n)$. Let $N = 2^n + 1$ and let $\X=L\cup R\cup P\cup D$  where
\begin{align*}
    L &= \{ a^l, b_i^l, c^l_i \mid 0 \leq i < N\}\\
    R &= \{ a^r, b_i^r, c^r_i \mid 0 \leq i < N \}\\
    P &= \{p, p'\}\\
    D &= \{d_1, d_2, d_3\}
\end{align*}
Keep an eye on Figure~\ref{fig:domainsSn}. 
We define a partial binary function $\bullet$ over $\X$ as follows.   Over $L\times R$ let
$$\begin{array}{ccc}
    b^l_i \bullet c^r_i = p' & b^l_i \bullet b^r_{i + 1} = p' & b^l_{i + 1} \bullet c^r_i = p'   \\
    c^l_i \bullet b^r_i = p' & c^l_i \bullet c^r_{i + 1} = p' & c^l_{i + 1} \bullet b^r_i = p'
\end{array}$$
for all $0 \leq i < N$ where addition of indices is modulo $N$, and for any other pair of elements $l \in L, r \in R$ we let $l\bullet r = p$.   For pairs not in $L\times R$, let
\[
l\bullet d_2=d_1,\;\;\; r\bullet d_3=d_2,\;\;\; p\bullet d_3=p'\bullet d_3=d_1\]
for $l\in L,\; r\in R$, and let $\bullet$ be undefined on all other pairs.   For any subsets $S, T$ of $\s_n$ let
\[S\bullet T = \{s \bullet t \mid s \in S,\; t \in T, s \bullet t  \mbox{ is defined}\}.\]

Now we define for all $A \subseteq \X$ the closure (denoted $\widehat{A}$) as the limit of repeating the following steps:
\begin{enumerate}
 \item \label{en:du} if $|\{d_1, d_2, d_3\} \cap A| > 1$, set $A = \emptyset$
    \item \label{en:d1} if  $A\cap(L \cup P)\neq\emptyset $,  add $d_1$ to $A$
    \item \label{en:d2} if $A \cap R\neq\emptyset$, add $d_2$ to $A$
    \item \label{en:nd} for each $0 \leq i < N$
        \begin{enumerate}
            \item  if $\set{b^l_i, c_i^l} \subseteq A$, add $a^l$ to $A$
            \item if $\set{b^r_i, c_i^r}\subseteq A$, add $a^r$ to $A$
        \end{enumerate}
           
\end{enumerate}
since each iteration either expands $A$ or replaces it by $\emptyset$, this limit is well-defined.    
Let the underlying set of $\s_n$  be
$$\{\widehat{A} : A \subseteq \X\}.$$
To define $+,\circ$, let
\begin{align*}
    S+\emptyset&=\emptyset+S=\emptyset\\
    S + T &= \widehat{S \cup T}\;\; (S, T\neq\emptyset)\\
    S \circ T &= \widehat{S \bullet T}
\end{align*}
for all $S, T\in\c S_n$.

The top element of this join-semilattice is $\emptyset$ and steps \eqref{en:du}, \eqref{en:d1}, \eqref{en:d2} of $\widehat{\;}$ ensure that each non-empty set $\widehat{S}$ is contained in exactly one of $L\cup P\cup\set{d_1}, \; R\cup\set{d_2}$ or $\set{d_3}$ and includes a unique \emph{domain} element,  which we denote as $\delta(\widehat{S})\in\set{d_1, d_2, d_3}$ (see Figure~\ref{fig:domainsSn}).

The $+$ defined for the structures above is non-distributive, i.e. it does not hold for every $a,b,c$ that if $a \leq b+c$ there exist some $b',c'$ such that $b' \leq b, \;c' \leq c$ and $b'+c'=a$ \/ (to see this, take  $b=\set{d_1, b^l_i},\; c=\set{d_1, c^l_i},\;b+c=\set{d_1, b^l_i, c^l_i, a^l},\; a=\set{d_1, a^l}$, for any $i<N$ and similar with $(r, 2)$ in place of $(l, 1)$).

Since the structures are non-distributive, we cannot assume that irreducible elements are prime.   The irreducible elements are singletons $\set{d_1},\;\set{d_2},\;\set{d_3}$ and doubletons $\set{d_1, s},\;\set{d_2, r}$ where $s\in L\cup P,\; r\in R$.  All of these are prime, except $\set{d_1, a^l}$ and $\set{d_2, a^r}$ which fail to be prime, as we saw.

Our argument, as well as that in \cite{andreka2011axiomatizability}, for non-finite axiomatisability heavily relies on the failure of distributivity. However, our structures differ from those in \cite{andreka2011axiomatizability} in the following aspect.

\begin{lemma}
\label{lem:localTot}
Suppose $\N$ occurs in a play of the game, $S\in\top(x, y),\; T\in\bot(x)$
where $\emptyset \neq  S\subseteq T \in \s_n$. Then $\forall$ can win the game in at most two moves.
\end{lemma}

\begin{proof}
We may assume $S\neq T$ else $\N$ is already inconsistent.    Let $\delta(S)=\delta(T)=d_i$, where $i=1, 2$ or $3$.    Since $T\supsetneq\set{d_i}$ we know $i\neq 3$ and $T$ has non-empty intersection with either $L, R$ or $P$.    Hence, there is $j>i$ such that $T\circ \set{d_j}=\set{d_i}$.

 $\forall$ can play a choice move $S+\set{d_i}=S$ to force  an outgoing edge $(x, z)$ with $\set{d_i}\in\top(x, z)$.  But then, a witness move for $T\circ\set{d_j}=\set{d_i}$ forces an edge $(x, w)$ with $T\in\top(x, w)$, yielding an inconsistent network.
\end{proof}

We now have all the tools to show the following.

\begin{lemma}
\label{lem:unrep}
$\s_n$ is not $(\sqcup, ;)$-representable for any $1 < n < \omega$.
\end{lemma}

\begin{proof}
We show that $\forall$ has a winning strategy in $\Gamma_{2N+1}(\s_n)$. Assume $\forall$ picks $\{p', d_1\} \neq \{p,p',d_1\}$. If $\exists$ returns the network where $\{p',d_1\}\in \top(x_0, y_0)$ and $s_\bot = \{p,p',d_1\}$, $\forall$ wins by Lemma~\ref{lem:localTot}.

Now, let us have a look at the case where $\exists$ puts $\{p,p',d_1\}$ in  $\top(x_0, y_0)$ and $s_\bot=\set{p', d}$. When faced with the choice move $\{p,p',d_1\} = \{p,d_1\} + \{p',d_1\}$ over $x_0, y_0$, she must put $\{p,d_1\} \in \top(x_0, y_0)$ as $s_\bot = \{p',d_1\}$. $\forall$ may request a witness over $x_0,y_0$ for $\{p,d_1\} = \{a^l,d_1\} \circ \set{a^r, d_2}$, let us call the node $\exists$ chooses to witness this composition $z$.

If $\forall$ plays a join move $\set{a^l, d_1}+\set{b^l_i,c^l_i,a^l, d_1}$ over $(x_0, z)$, by Lemma~\ref{lem:localTot}, \/$\exists$ will have to add $\set{b^l_i,c^l_i,a^l, d_1}$ to $\top(x_0, z)$ or lose the game, and similarly she can be forced to add $\set{b^r_i,c^r_i,a^r, d_2}$ to $\top(z, y_0)$. 
Observe that   $\set{b^l_i,c^l_i,a^l, d_1}=\set{b^l_i, d_1}+\set{c^l_i, d_1}$, so
over a series of $2N$ choice moves,  she must  add either $\{b^l_i,d_1\}$ or $\{c^l_i,d_1\}$ to $\top(x_0,z)$  and either $\{b^r_i,d_2\}$ or $\{c^r_i,d_2\}$ to $\top(z,y_0)$, for $i<N$.

Without loss, assume she picks $\{b_0^l, d_1\}$ for $(x_0, z)$. To avoid a composition resulting in $\{p', d_1\}$ she must also pick $\{b_0^2, d_1\}$ for $(z, y_0)$. Again to avoid the undesired composition, she must also pick $\{c_1^l, d_1\}, \{c_1^r, d_2\}$ for $\top(x_0, z), \top(z, y_0)$ respectively. More generally, she must add $\{c_i^l, d_1\}, \{c_i^r, d_2\}$ if $i$ is odd and $\{b_i^l, d_1\}, \{b_i^r, d_2\}$ if $i$ is even. Since $N-1$ is even, she must choose $\{b_{N-1}^l, d_1\}, \{b^r_{N-1}, d_2\}$ and this results in a composition  $\{b_{N-1}^l, d_1\}\circ \{b_0^r, d_2\}=\set{p', d_1}=s_\bot$, so the network is  inconsistent.
\end{proof}

\begin{lemma}
\label{lem:winIfGoodPair}
$\exists$ can win the game $\Gamma_\omega(\s_n)$ if $\forall$ does not use the initialisation pair $\{p', d_1\} \neq \{p,p',d_1\}$.
\end{lemma}

\begin{figure}
    \centering
    \begin{tikzpicture}
        \node[draw, circle] (x) at (0,0) {$x$};
        \node[draw, circle] (y) at (0,3) {$y$};
        \node[draw, circle] (z) at (0,-5) {$z$};
        \node[draw, circle] (u) at (3.5,-1.2) {$u$};
        \node[draw, circle] (w) at (2,1.5) {$w_{l,r}$};
        \node[draw, circle] (v) at (2,-3.5) {$v_{\rho,\rho'}$};
        \node[draw, circle] (f) at (6,0) {$\times$};
        
        \draw[dashed] (.5, .2) rectangle (3,2.8);
        \node[draw, dashed] (lt) at (3, 4) {$l \in L, r \in R: l \bullet r = p'$};
        \path (3, 2.5) edge[dashed] (lt);
        
        \path (y) edge[bend left=50, -->-] node[right, pos=0.8]{$d_3^\uparrow$} (f);
        \path (w) edge[bend left, ->-] node[above]{$d_2^\uparrow$} (f);
        \path (w) edge[->-] node[above right, pos=0.45]{$r^\uparrow$} (y);
        \path (x) edge[->-] node[below right, pos=0.65]{$l^\uparrow$} (w);
        
        \draw[dashed] (.5, -2.2) rectangle (3,-4.8);
        \node[draw, dashed] (lb) at (3, -6) {$\rho, \rho'\subseteq\set{i:i<N}$};
        \path (3, -4.8) edge[dashed] (lb);
        
        \path (z) edge[bend right=60, -->-] node[right, pos=0.8]{$d_3^\uparrow$} (f);
        \path (v) edge[bend right, ->-] node[below]{$d_2^\uparrow$} (f);
        \path (v) edge[->-] node[below right,pos=.4]{${B^r(\rho')}$} (z);
        \path (x) edge[-->-] node[left, pos=0.9]{${B^l(\rho)}$} (v);
        
        \path (x) edge[->-]node[left]{$p'^\uparrow$} (y);
        \path (x) edge[->-]node[left]{$p'^\uparrow \cup p^\uparrow$} (z);
        \path (x) edge[-->-]node[above, pos=.8]{$d_1^\uparrow$} (f);
        
        \path (x) edge[->-]node[above right]{$p'^\uparrow \cup p^\uparrow$} (u);
        \path (u) edge[->-]node[below right]{$d_3^\uparrow$} (f);
        \path (w) edge[->-, bend left]node[right]{$d_2^\Uparrow$} (u);
        \path (v) edge[-->-]node[left, pos=0.8]{$d_2^\Uparrow$} (u);
    \end{tikzpicture}
    \caption{A saturated network}
    \label{fig:win_net}
\end{figure}
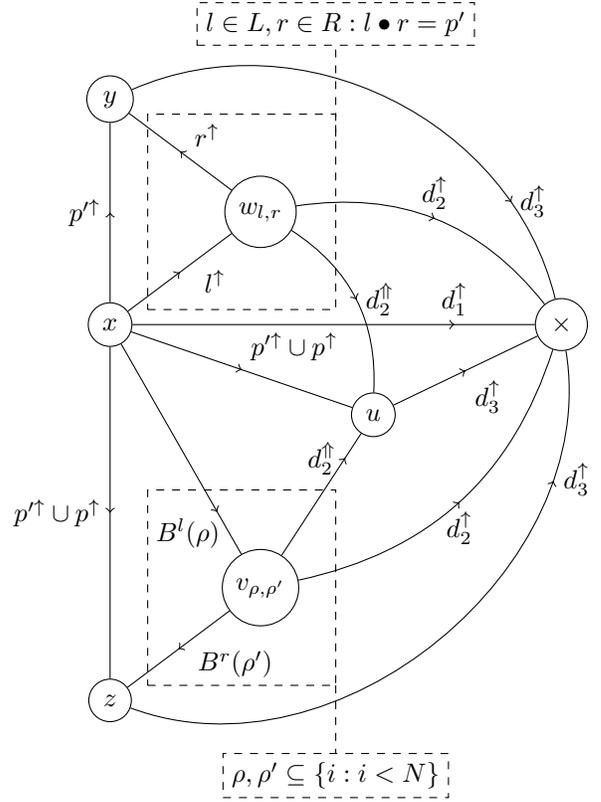

\begin{proof}
We show this by explicitly defining a saturated network $\N$,  discriminating all pairs except $\set{p', d_1}\neq\set{p, p', d_1}$ (see figure~\ref{fig:win_net}).  Since $\exists$ is not required to play conservatively, she may play $\N$ in the initial round, by selecting $x_0, y_0\in\N,\;s_\bot\in\s_n$ appropriately.

First, some  notation and a definition.  For any $S\subseteq \X$ let
\begin{align*}
    S^\uparrow&=\set{T\in \s_n: S\subseteq T}\\
    S^\Uparrow&= S^\uparrow\setminus \set S
    \end{align*}
and for 
$s\in \X$ write  $s^\uparrow, s^\Uparrow$ for $\set{s}^\uparrow, \set{s}^\Uparrow$.

Let $S\subseteq \s_n$ be \emph{upward closed}, i.e. 
\[ (A\in S \wedge A\subseteq B\in\s_n)\rightarrow B\in S\]
In view of Lemma~\ref{lem:localTot} we may assume that $\exists$ chooses upward closed sets for each label $\top(x, y)$.  
We say that $S$ is  \emph{prime} if it is upward closed and for all $A, B\in\s_n$ we have $(A+B)\in S \rightarrow(A\in S\vee B\in S)$.   The edge labels $\top(x, y)$ of a saturated network must be prime.
 
Note that $s^\uparrow$ is prime, for $s\in \X\setminus\set{a^l, a^r}$, but $(a^l)^\uparrow, (a^r)^\uparrow$ are not.  Also note that $d_2^\Uparrow=\set{\set{d_2}\cup R_0:\emptyset\neq R_0\subseteq R}$ is prime.   A prime set including $S\ni a^l$ must include an element including either $b^l_i$ or $c^l_i$, for each $i<N$.  
So, for any $\rho\subseteq\set{i:i<N}$, let 
\begin{align*}
    B^l(\rho)&=(a^l)^\uparrow\cup
\bigcup_{i\in\rho}(b^l_i)^\uparrow\cup\bigcup_{i<N,\; i\not\in\rho} (c^l_i)^\uparrow\\
B^r(\rho)&= (a^r)^\uparrow\cup\bigcup_{i\in\rho}(b^r_i)^\uparrow\cup\bigcup_{i<N,\; i\not\in\rho}(c^r_i)^\uparrow
    \end{align*}
 and observe that these sets are prime.

The set of nodes $N$ of $\N$ is defined to be 
\begin{align*}
         & \{x,y,z,u,\times\} \\
    \cup & \{w_{l,r} : l \in L,\;  r \in R,\;  l \bullet r = p'\}\\
    \cup & \{v_{\rho, \rho'}: \rho, \rho'\subseteq\set{i:i<N}\}
\end{align*}
Edge labels between nodes are as shown in Figure~\ref{fig:win_net}, all edges not shown have empty edge labels.   Let the index of $x$ be 1, the indices of $v_{\rho, \rho'}$ and $w_{l, r}$ be 2 and the indices of $u, y, z$ be $3$ ($\times$ has no index).  For the node labelling of any $q \neq \times $ with index $i \in\set{1,2,3}$,  let $\bot(q)=\set\emptyset\cup \set{S\in\s_n:\delta(S) \neq d_i}$ and let $\bot(\times)=\s_n$.  

It can be checked exhaustively that for every pair but $\{p', d_1\} \neq \{p,p',d_1\}$, there exists a $x, y\in\N$ where $\top(x, y)$ includes one but not the other of the pair, for example if $B, C, B', C'\subseteq\set{i:i<N},\; (B, C)\neq (B', C')$, so $\set{d_1}\cup \set{b_i:i\in B}\cup\set{c_i:i\in C} \neq \set{d_1}\cup \set{b_i:i\in B'}\cup\set{c_i:i\in C'} $, this pair is  discriminated on $(x, v_{\rho, \rho'})$ provided $
\rho\subseteq B,\;\rho\not\subseteq B'$, or if $\rho\cup C=\set{i:i<N},\;\rho\cup C'\neq\set{i:i<N}$, or the other way round.  It can also be checked that  the network is saturated.  
\end{proof}

An $\s$-network $\N$ is \emph{prime} if for all $x, y\in\N$,\/   $\top(x, y)$ is a prime subset of $\s_n$.  A network $\N$ is \emph{indexed} if each node $x\in\N$ has an index $\iota(x)\in\set{1, 2, 3}$ such that for all $x, y\in\N,\;(S\in\top(x, y)\rightarrow \delta(S)=d_{\iota(x)})$.  

\begin{lemma}
\label{lem:winN}
Let $\N$ be a closed, prime, consistent, indexed network.  Then $\exists$ has a winning strategy in $\Gamma_n(\s_n, \N)$.

\end{lemma}

\begin{proof}   Assume $\N$ satisfies the conditions.   Extend the labelling of $\N$  so that
\[ \bot(x)=\set{S\in\s_n: \delta(S)\neq d_{\iota(x)}}\]

The network remains prime, closed, indexed and consistent, by Lemma~\ref{lem:localTot}, and  this ensures that all join moves are trivial.
Note that  by closure of $\N$, all composition moves are trivial,  and since edge labels are prime sets,  choice moves  are also trivial.   

That leaves witness moves.  If $\forall$ plays a witness move $(x, y, \alpha, \beta)$ where $\alpha\circ\beta\in\top(x, y)$, by the additive definition of $\circ$ and since $\top(x, y)$ is prime, we can find join irreducibles $\alpha_0\subseteq\alpha,\;\beta_0\subseteq\beta$ where $\alpha_0\circ\beta_0\in\top(x, y)$.  

Let $\delta(\beta_0)=d_i$ where $i\in\set{2,3}$.
She lets $\top(x, z)=\alpha_0^\uparrow,\;\top(z, y)=\beta_0^\uparrow$  where $z$ is a new node  with index $i$, and extends the network to $z$ by $\top(x, z)=\alpha_0^\uparrow,\;\top(z, y)=\beta_0^\uparrow,\;\bot(z)=\set{S\in\s_n:\delta(S)\neq d_i}$.  Since $\alpha_0\circ\beta_0\in\top(x, y)$ it follows that  $\alpha_0\subseteq\set{d_1}\cup L$ and  $\beta\subseteq \set{d_2}\cup R$.  Hence the resulting network will be closed.   
If $\alpha_0, \beta_0$ are primes she can play  a prime network and from there she can win the game of length $n-1$ by the induction hypothesis.  So we may assume either $\alpha_0$ or $\beta_0$ is irreducible but not prime, and since $\alpha_0\circ\beta_0\in\top(x, y)$, either $\alpha_0=a^l$ or $\beta_0=a^r$, without loss assume the former $\alpha_0=a^l$, and $\top(x, y)=\set{p}^\uparrow$.  If $\beta_0$ is prime, then choose some prime set $\pi$, such that $\pi\circ \beta_0^\uparrow=p^\uparrow$ and let $\top(x, z)=\pi$.

Finally, suppose neither $\alpha_0$ nor $\beta_0$ is prime, so $\alpha_0=a^l,\;\beta_0=a^r$.  We must show that $\exists$ can survive $n-1$ choice moves, the remainder of the game.  Inductively, she maintains a contiguous set $\Delta\subseteq\set{i:i<N}$ whose complement in $\set{i:i<N}$ has size more than $2^k$ where $k$ is the number of rounds remaining,  and a partition of $\Delta$ into two  successor-free sets $\Delta_b\stackrel{\bullet}{\cup}\Delta_c=\Delta$, for parity.  Initially $\Delta=\emptyset$.  She ensures that $S\in\top(x, z)\rightarrow S\subseteq\set{d_1, a^l}\cup\set{b^l_i:i\in\Delta_b}\cup\set{c^l_i:i\in\Delta_c}$, and the similar for $T\in\top(z, y)$.  This ensures $S\circ T\in p^\uparrow$ for $S\in\top(x, z),\; T\in\top(z, y)$ which proves closure of the network.   To maintain this induction hypothesis, when $\forall$ plays a choice move $\set{d_1, b^l_i}+\set{d_1, c^l_i} \in\top(x, y)$ she extends $\Delta$ to include $i$, most economically, and this ensures that the complement in $\set{i:i<N}$ will be at least half its previous size, which maintains the induction hypothesis.  This completes the winning strategy.  
\end{proof}

\begin{lemma}\label{lem:ws}
$\exists$ has a winning strategy in $\Gamma_n(\s_n)$.
\end{lemma}
\begin{proof}
In the initial round let $\forall$ play $a\neq b\in\s_n$.  
There is a prime $\pi\in\s_n$ such that $\pi\subseteq a,\;\pi\not\subseteq b$ or $\pi\not\subseteq a,\; b\subseteq\pi$, without loss assume the former.  $\exists$ plays the prime network $\N_0$ with nodes $x_0, y_0$ and lets $\top(x_0, y_0)=\pi^\uparrow$, otherwise $\top(x', y')$ is empty.   
She defines a suitable node index function $\iota$ from $\pi$, e.g. if $\pi=\set{d_1, b^l_i}$ she lets $\iota(x_0)=1,\;\iota(y_0)=2$.    Then $\N_0$ is closed, prime, consistent and has a suitable index function. 
By Lemma~\ref{lem:winN} she can survive another $n$-rounds.
\end{proof}
We now have everything to claim

\begin{theorem}
$R(\sqcup, ;)$ is not finitely axiomatisable.
\end{theorem}

\begin{proof}
Suppose the class was axiomatised by a single first order formula $\psi$. Since the theory $\set{\sigma_n : n<\omega}$ axiomatises $R(\sqcup, ;)$, we know that $\set{\sigma_n : n<\omega} \cup \{\neg \psi\}$ is not satisfiable. 
Take any  finite subtheory, let $N$ be maximum such that $\sigma_N$ is included in the subtheory.   By Lemma~\ref{lem:unrep} the structure $\s_{N}$ is not representable, hence it is a model of $\neg\psi$ but by Lemmas~\ref{lem:ws} and \ref{lem:sig}, it  satisfies $\sigma_N$, hence it satisfies the whole finite subtheory.  By compactness we have reached a contradiction.

\end{proof}

\section{Semigroups with Demonic Meet}
\label{sec:meet}

In this section we look at semigroups with the demonic meet, denoted $(\cdot, \circ)$ in the abstract and $(\sqcap, ;)$ in the concrete signature. We define $\sqcap$ as the greatest lower bound with respect to $\sqsubseteq$.  We noted in the introduction that not every pair of binary relations has a common refinement, but this could be fixed by adjoining an extra point $\times$ to the base of the relations and replacing each relation $R$ by $R\cup\set{(x, \times):x\in \dom(R)}$.


Now, we slightly modify  the Point Algebra to show the following

\begin{theorem}
Finite representation property of finite structures fails in $R(\sqcap, ;)$
\end{theorem}

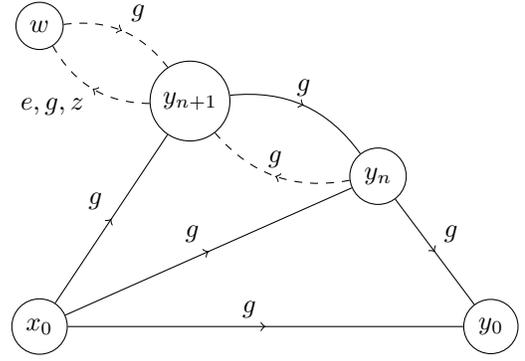
\begin{figure}
    \centering
    \begin{tikzpicture}
        \node[draw, circle] (x0) at (0,0){$x_0$};
        \node[draw, circle] (y0) at (6,0){$y_0$};
        \node[draw, circle] (yn) at (2,3){$y_{n+1}$};
        \node[draw, circle] (yi) at (4.5,2){$y_n$};
        \node[draw, circle] (w) at (0,4){$w$};
        
        \path (x0) edge[->-] node[above left] {$g$}(yn);
        \path (x0) edge[->-] node[above left] {$g$}(yi);
        \path (x0) edge[->-] node[above left] {$g$}(y0);
        \path (yi) edge[->-] node[above right] {$g$}(y0);
        \path (yn) edge[->-, bend left] node[above] {$g$}(yi);
        \path (yi) edge[->-, bend left, dashed] node[above] {$g$}(yn);
        \path (yn) edge[->-, bend left, dashed] node[below left] {$e,g,z$}(w);
        \path (w) edge[->-, bend left, dashed] node[above right] {$g$}(yn);
    \end{tikzpicture}
    \caption{Induction Case in Proof for NFRP $R(\sqcap, ;)$}
    \label{fig:indMeet}
\end{figure}

\begin{proof}
We explicitly define a finite algebra $\s$. The underlying set of the algebra is $\{z,e,g\}$ and the two binary operations $\cdot, \circ$  are defined by
\begin{center}
 \begin{tabular}{|c|ccc|}
    \hline
    $\cdot$ & $z$ & $e$ & $g$ \\
    \hline
    $z$ & $z$ & $z$ & $z$  \\
    $e$ & $z$ & $e$ & $z$  \\
    $g$ & $z$ & $z$ & $g$  \\ \hline
\end{tabular}\hspace{1cm}
\begin{tabular}{|c|ccc|}
    \hline
    $\circ$ & $z$ & $e$ & $g$ \\
    \hline
    $z$ & $z$ & $z$ & $z$  \\
    $e$ & $z$ & $e$ & $g$  \\
    $g$ & $z$ & $g$ & $g$  \\ \hline
\end{tabular}
\end{center}
See how this algebra is representable over the base $\mathbb{Q} \cup \{ \bot \}$ by representation $\theta$ where
\begin{align*}
    z^\theta &= \{(q, \bot) \mid q \in \mathbb{Q}\} \cup \{(\bot, \bot)\}\\
    e^\theta &= \{(q,q) \mid q \in \mathbb{Q}\} \cup z^\theta\\
    g^\theta &= \{(q,r) \mid q<r \in \mathbb{Q}\} \cup z^\theta
\end{align*}

Now we show that any representation $\theta$ has to be infinite.  Let us look at any representation of $\s$. There must  exist a pair of nodes $(x_0, y_0)$ that is either included in $z^\theta \setminus g^\theta$ or $g^\theta \setminus z^\theta$.

Suppose that $(x_0, y_0) \in z^\theta \setminus g^\theta$. Since $z = g \circ z$ it follows that $x_0 \in \dom(g^\theta)$. Because $z \cdot g = z$ and $x_0 \in \dom(g^\theta)$ we have $(x_0, y_0) \in g^\theta$, yielding a contradiction.

Therefore, there must exist $(x_0, y_0) \in g^\theta \setminus z^\theta$.   Suppose that $(y_0, y_0)\in g^\theta$.  Since $g=e\circ g$ there is $w$ where $(y_0, w)\in e^\theta, \; (w, y_0)\in g^\theta$.  But then, $(y_0, w)\in e^\theta\cap (g^\theta;e^\theta)=z^\theta$, so $(x_0, y_0)\in (g\circ z\circ g)^\theta=z^\theta$, contrary to assumption.  Thus $(y_0, y_0)\not\in g^\theta$. 

We now show by induction that there must exist distinct points $y_0, y_1, \ldots, y_n$,  such that 
\begin{enumerate}[label=\Roman*]
\item\label{ih:1}$i<j\leq n \rightarrow (y_j,y_i) \in g^\theta$ and
\item\label{ih:2} $i\leq n\rightarrow (y_i, y_i)\not\in g^\theta$.
\end{enumerate}  The base case, $n=0$, has been established.  

For the induction case, see Figure~\ref{fig:indMeet}.  Since $(x_0, y_n)\in g^\theta=(g\circ g)^\theta$, there must be a point $y_{n+1}$ such that $(x_0, y_{n+1})\in g^\theta,\; (y_{n+1}, y_n)\in g^\theta$.    If $y_{n+1}\in\set{y_0, \ldots, y_n}$ then $(y_n, y_n)\in g^\theta\cup (g\circ g)^\theta=g^\theta$, contradicting \eqref{ih:2}.  Hence $y_{n+1}$ is distinct from $\set{y_0, \ldots, y_n}$.  Using $g=g\circ g$ again, we deduce $(y_{n+1}, y_{i})\in (g\circ g)^\theta=g^\theta$ (for all $i\leq n$), establishing \eqref{ih:1} for $n+1$.  We cannot have $(y_{n+1}, y_{n+1})\in g^\theta$, else by $g=e\circ g$ there would be $w$ with $(y_{n+1}, w)\in e^\theta, \; (w, y_{n+1})\in g^\theta$, so $(y_{n+1}, w)\in (e\cdot g)^\theta=z^\theta$, as before, hence $(x_0, y_0)\in (g\circ z\circ g)^\theta=z^\theta$, contrary to assumption.  This establishes \eqref{ih:2} for $n+1$.

From above induction we conclude that for every $n < \omega$, the base has at least $n$ distinct points. Thus any representation of $\s$ is over an infinite base.
\end{proof}

\section{Semigroups with Demonic Lattice}
\label{sec:lattice}

\begin{lemma}\label{lem:zz}
Let $\c A=(A, \cdot, +, \circ)$  be a lattice with a binary operator $\circ$, with least element ${\bf 0}$ such that $a\circ {\bf 0} ={\bf 0}$ for all $a\in\c A$.    Then $\c A\in R( \cap, \cup, ;)$ if and only if $\c A\in R(\sqcap,\sqcup, ;)$, and $\c A$ has a $(\cap, \cup, ;)$-representation on a finite  base if and only if it has a $(\sqcap,\sqcup, ;)$-representation on a finite base.

\end{lemma}
\begin{proof}
Let $\theta$ be a $(\cap, \cup, ;)$-representation of $\c A$ over base $X$.  Define $\psi:\c A\rightarrow\wp((X\cup\set\bot)\times (X\cup\set\bot))$ by $a^\psi=a^\theta \cup\set{(x, \bot):x\in X\cup\set\bot}$ where $\bot\not\in X$, so $\psi$ is a $(\cap, \cup, ;)$-representation, representing each $a\in\c A$ as a left-total binary relation over $X\cup\set\bot$.  Since demonic and angelic operators agree over left-total relations, it follows that $\psi$ is a $(\sqcap, \sqcup, ;)$-representation of $\c A$ (and the base is finite if $X$ is). 

Conversely, suppose $\phi$ is a $(\sqcap, \sqcup, ;)$-representation of $\c A$ over $X$.    Since ${\bf 0}\leq a$ in $\c A$ we have $\phi({\bf 0})\sqsubseteq \phi(a)$ so $\dom(\phi({\bf 0}))\supseteq \dom(\phi(a))$.  Since ${\bf 0}=a \circ {\bf 0}$ we get $\dom(\phi({\bf 0}))\subseteq \dom(\phi(a))$, hence we get $\dom(\phi(a))=\dom(\phi({\bf 0}))$ for all $a\in\c A$.    Since $\dom(\phi(a))$ is constant, demonic and angelic operators agree, so $\phi$ is a $(\cap, \cup, ;)$-representation of $\c A$ (whether its base is finite or not).
\end{proof}

\begin{theorem}
The $(\sqcap,\sqcup, ;)$-representation problem and the finite $(\sqcap,\sqcup, ;)$-representation problem are both undecidable for finite $(\cdot, +, \circ)$-structures.

\end{theorem}
\begin{proof}
A finite partial group $*$ is a total, binary, surjective map $*:P\times P\rightarrow A$ for finite sets $P, A$.  It is a yes-instance of the  group embedding problem if there is a group $G$ and an injection from $A$ into $G$ preserving all defined products, a no-instance otherwise.
It is a yes-instance of the finite group embedding problem if there is a finite group $G$ and an embedding from $A$ into $G$. 
Both these problems are known to be undecidable \cite{slo:81,KS:95}.  

In \cite{HJ:undec} a Boolean monoid $\M(*)$ is constructed from a partial group $*$ (in \cite{HJ:undec}, the definition of partial group is slightly different,   the partial group is called ${\bf A}$ and the Boolean monoid is $\M({\bf A})$, but our construction and notation here are equivalent).   \cite[Proposition~5.1]{HJ:undec} proves that  $*$ is a yes-instance of the  group embedding problem  iff $\M(*)\in R(\cap, \cup,1', ;)$, and  $*$ is a yes-instance of the  finite group embedding problem iff $\M(*)$ has a $(\cap, \cup,1', ;)$-representation on a finite base. 

Now $\M(*)$ is a monoid with an identity $1'$, and a $(\cap, \cup, ;)$ representation need not represent $1'$ as the true identity over its base.  However,   \cite[Lemma~3.1]{neuzerling2016undecidability}  shows that we may quotient the base of such a representation by the binary relation consisting of all pairs $(x, y)$ of points where either $x=y$ of  $(x, x), (x, y), (y, x)$ and $(y, y)$ are in the representation of $1'$ to obtain a $(\cap, \cup, 1',;)$-representation of $\M(*)$.   Hence $*$ is a yes-instance of the (finite) group embedding problem iff $\M(*)$ has a $(\cap, \cup, ;)$-representation (on a finite base).

 By  Lemma~\ref{lem:zz} (noting that $\M(*)$ has a zero), this statement is  equivalent  to $\M(*)$ being  $(\sqcap, \sqcup, ;)$-representable (respectively,  finitely $(\sqcap, \sqcup, ;)$-representable).  The Theorem follows.
\end{proof}

\begin{corollary}
$R(\sqcap, \sqcup, ;)$ cannot be defined by finitely many axioms.  There are finite algebras in $R(\sqcap,\sqcup, ;)$ with no finite representations.

\end{corollary}
\begin{proof}
Finite axiomatisabillity implies decidability of the representation problem for finite algebras, false by the theorem above. Since $R(\sqcap, \sqcup, ;)$ has a recursively enumerable axiomatisation, its finite members are co-recursively enumerable (we may enumerate the non-representable finite algebras by checking if they fail any of the axioms using a fair schedule).  If the representation class had the finite representation property then the finite, representable algebras would also be recursively enumerable, hence recursive,  contradicting the previous theorem.
\end{proof}

\section{Conclusion and Problems}
Although the combination of demonic lattices with ordinary composition provides us with an intuitive relational way of modelling total correctness and termination of nondeterministic machines, the results in the area suggest computational problems. 

One significant exception is the recent result that finite structures in $R(\sqsubseteq, ;)$ have a finite representation property with quadratic upper bound on its size, despite the class not being finitely axiomatisable \cite{hirsch2020finite}. This implies that representability is not only decidable for finite structures, but also provides a nondeterministic polynomial upper bound on the complexity of checking representability.

Furthermore, although we have shown that $R(\sqcup, ;)$ is not finitely axiomatisable, we had to do so utilising structures with nondistributive semilattices and the argument does not easily translate to the subclass of structures with distributive semilattices. It has been shown \cite{andreka2011axiomatizability} that its angelic counterpart is both finitely axiomatisable and has the finite representation property. Thus an interesting problem that remains open

\begin{problem}
Is the subclass of $(\sqcup, ;)$-representable $(+,\circ)$-structures with distributive $+$ finitely axiomatisable? Do its finite members have the finite representation property?
\end{problem}

Furthermore, some good behaviour may be found in answers to the following two open questions.

\begin{problem}
Do the finite members of $R(\sqcup, ;)$ have the finite representation property?
\end{problem}

\begin{problem}
Is $R(\sqcap, ;)$ finitely axiomatisable?
\end{problem}

The answers for the former in the angelic form is also open and $R(\cap, ;)$ is known to be finitely axiomatisable.

\section*{Acknowledgement}
The authors would like to thank Mr. D. Rogozin for insightful conversations and making them aware of some existing results

\bibliographystyle{IEEEtran}
\bibliography{ref}

\end{document}